\documentclass[prc,amssymb, amsmath, nofootinbib,tightenlines,
nobibnotes, showpacs,showkeys,aps, prl,twocolumn]{revtex4-1}
\usepackage{graphicx}
\begin{document}
%
%\title{Thermodynamic consistency of the relativistic quantum form of the Tsallis distribution.}
\title{The Tsallis Distribution and Transverse Momentum Distributions in High-Energy Physics.}
\author{J. Cleymans and D. Worku}
\affiliation{UCT-CERN Research Centre and Department of Physics, University of Cape Town, 
Rondebosch 7701, South Africa}
\date{\today}
\begin{abstract}
The Tsallis distribution has been used recently to fit the 
transverse momentum distributions of identified particles
by the STAR collaboration~\cite{Abelev:2006cs} at 
the Relativistic Heavy Ion Collider 
and by the ALICE~\cite{Aamodt:2011zj} and CMS~\cite{Khachatryan:2011tm} 
collaborations at the Large Hadron Collider.
Theoretical issues are clarified concerning the thermodynamic 
consistency of the Tsallis distribution 
in the particular case of 
relativistic high energy quantum distributions.    
An improved  form is proposed for describing the transverse 
momentum distribution and fits are presented together with estimates of
the parameter $q$ and the temperature $T$.
\end{abstract}
\maketitle 
\section{\label{secIntroduction}Introduction}
The Tsallis distribution has gained prominence recently  
in high energy physics with very high quality fits
of the transverse momentum distributions  made
by the STAR collaboration~\cite{Abelev:2006cs} at the 
Relativistic Heavy Ion Collider 
and by the ALICE~\cite{Aamodt:2011zj} and CMS~\cite{Khachatryan:2011tm}
 collaborations at the Large Hadron Collider.

In the literature there exists more than one version of the Tsallis 
distribution~\cite{Tsallis:1987eu,Tsallis:1998ab}
and we would like to investigate in this paper  one version which 
 we consider suited for
describing results in high energy particle physics.
Our main guiding criterium will be thermodynamic consistency which
has not always been implemented 
correctly (see e.g.~\cite{Pereira:2007hp,Pereira:2009ja,Conroy:2010wt}).
The explicit form which we use for the 
transverse momentum distribution in relativistic heavy ion collisions
 is: 
\begin{eqnarray}
&&\frac{dN}{dp_T~dy} = \nonumber\\
&&gV\frac{p_Tm_T\cosh y}{(2\pi)^2}
\left[1+(q-1)\frac{m_T\cosh y -\mu}{T}\right]^{q/(1-q)},
\label{Tsallis-B}
\end{eqnarray}
where $p_T$ and $m_T$ are the transverse momentum and mass respectively, $y$
is the rapidity, $T$ and $\mu$ are the temperature and the chemical potential,
the other variables are defined below.
In the limit where the parameter $q$ goes to 1 this reproduces 
the standard Boltzmann distribution:
\begin{eqnarray}
&&\lim_{q\rightarrow 1}\frac{dN}{dp_T~dy} = \nonumber\\
&&gV\frac{p_Tm_T\cosh y}{(2\pi)^2}
\exp\left(-\frac{m_T\cosh y -\mu}{T}\right).
\label{boltzmann}
\end{eqnarray}
In order to distinguish Eq.~\eqref{Tsallis-B} from the form used by 
the ALICE and CMS collaborations~\cite{Aamodt:2011zj, Khachatryan:2011tm} 
we will
refer to Eq.~\eqref{Tsallis-B} as the Tsallis-B  parameterization. 
Note in particular the extra power of $q$ on the right hand side.
The motivation for preferring this form is presented in detail 
in the rest of this paper, see
in particular section VII.

Thermal models have been successful in describing particle yields
at different beam 
energies~\cite{Cleymans:2005xv,Andronic:2005yp,Becattini:2005xt}, especially 
in heavy ion collisions.
These models assume the formation of a system which is in  
thermal and chemical equilibrium in the hadronic phase and are
characterized by a set of thermodynamic  variables for the
hadronic phase, most important among these are 
the chemical freeze-out temperature and baryon chemical potential. 
The deconfined period of the time evolution
dominated by quarks and gluons remains hidden: full equilibration generally
washes out and destroys large amounts of information about the early 
deconfined phase. 

While the description 
of integrated particle yields is reasonably successful, more detailed 
descriptions, especially of the transverse and longitudinal
momentum distributions call for additional dynamics.
The transverse momentum distribution is often described by a combination
of transverse flow and a thermodynamical statistical distribution. 
With the Tsallis distribution this superposition is not used 
and a very good fit can be obtained using  the additional parameter $q$
which describes the deviation from a Boltzmann distribution. 
In the limit where $q\rightarrow 1$ one recovers the standard
statistical Boltzmann distribution. Whether 
this is ultimately the correct description or not remains to be seen.
This paper is a contribution to the understanding of the use of the 
Tsallis distribution in high energy collisions. It is not meant as 
giving a final answer to the correct dynamical theory
of heavy ion collisions.\\
In the next section we review the derivation of the 
Tsallis distribution by emphasizing the quantum statistical form and  the
thermodynamic consistency. 
%The STAR~\cite{Abelev:2006cs}, ALICE~\cite{Aamodt:2011zj} 
%and CMS~\cite{Khachatryan:2011tm}
%collaborations have presented results
%on the transverse momentum distribution for identified particles. 
%
%
%
\section{Tsallis Distribution for Particle Multiplicities.}
%
%\subsection{Relation  between the Boltzmann and Tsallis distributions}
Several generalizations of the standard Fermi-Dirac distribution
\begin{equation}
n^{FD}(E) \equiv \frac{1}{1+\exp\left(\frac{E-\mu}{T}\right)}  .
\label{fd}
\end{equation}
to a Tsallis form 
have been proposed in the literature, some of these 
have been shown not to be thermodynamically consistent. 
In the following we use the Tsallis form of Fermi-Dirac distribution proposed 
in~\cite{turkey1,Pennini1995309,Teweldeberhan:2005wq,Conroy:2008uy,Conroy:2010wt}  
which uses 
\begin{equation}
n^{FD}_T(E) \equiv 
\frac{1}{1+\exp_q\left(\frac{E-\mu}{T}\right)}  .
\label{tsallis-fd}
\end{equation}
where the function $\exp_q(x)$ is defined as
\begin{equation}
\exp_q(x) \equiv \left\{
\begin{array}{l l}
\left[1+(q-1)x\right]^{1/(q-1)}&~~\mathrm{if}~~~x > 0 \\
\left[1+(1-q)x\right]^{1/(1-q)}&~~\mathrm{if}~~~x \leq 0 \\
\end{array} \right.
\label{tsallis-fd1}
\end{equation}
and, in the limit where $q \rightarrow 1$ reduces to the standard exponential:
$$
\lim_{q\rightarrow 1}\exp_q(x)\rightarrow \exp(x)
$$
The form given in Eqs.~\eqref{tsallis-fd} and \eqref{tsallis-fd1} will be 
referred to as the Tsallis-FD distribution. The Bose-Einstein version
(given below) will be referred to as the Tsallis-BE 
distribution~\cite{Chen200265} 
while the Boltzmann approximation will be referred to 
as Tsallis-B distribution. It should be noted that variations of the 
above have been presented previously in the literature. 
These will not be considered in this paper.

As is well-known, all forms of the   Tsallis distribution introduce a  
new parameter $q$. In practice this
parameter is always close to 1, e.g. in the 
results obtained by the ALICE and CMS collaborations
typical values for the parameter $q$ can be obtained  
from fits to the transverse momentum distribution for identified 
charged particles ~\cite{Aamodt:2011zj} and are close to the value 1.1 (see 
below).
The value of $q$ should thus be considered as never being far from 1, 
deviating from it by  20\% at most. An analysis of the composition of 
final state particles leads to a similar result~\cite{Cleymans:2008mt}
for the parameter $q$.

\begin{figure}
\begin{center}
\includegraphics[width=0.4\textwidth,height=10cm]{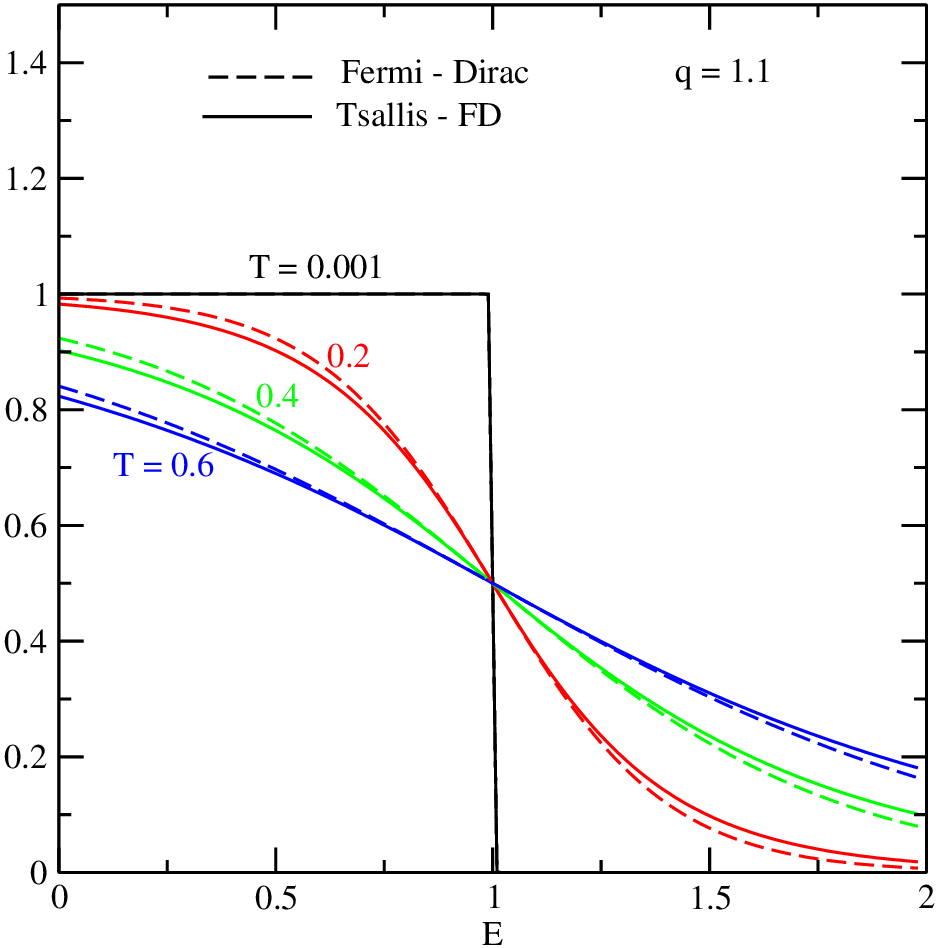}
\caption{Comparison between the Fermi-Dirac and Tsallis-FD distributions 
as a function of the energy $E$, keeping the Tsallis parameter $q$ fixed, for 
various values of the temperature $T$. 
The chemical potential is kept equal to one in all curves, the units
are arbitrary.}
\label{compare-tsallis-fd}
\end{center}
\end{figure}

In the limit where $q\rightarrow 1$ the two forms coincide. 
Numerically the difference is small, as shown in Fig.~(\ref{compare-tsallis-fd})
 for a 
value of $q = 1.1$. 

The Boltzmann approximation leads to the 
result~\cite{Tsallis:1987eu,Tsallis:1998ab}
\begin{equation}
n_T^B(E) = \left[ 1 + (q-1) \frac{E-\mu}{T}\right]^{-\frac{1}{q-1}} .
\label{tsallis}
\end{equation}
Note that we do not use the normalized $q$-probabilities which have been 
proposed in Ref.~\cite{Tsallis:1998ab} since we use here mean occupation numbers
which do not need to be normalized.  
In the limit where 
$q\rightarrow 1$ all distributions coincide with
the standard statistical distributions:
\begin{eqnarray}
\lim_{q\rightarrow 1} n_T^B(E)    &=& n^B(E), \\
\lim_{q\rightarrow 1} n_T^{FD}(E) &=& n^{FD}(E), \\
\lim_{q\rightarrow 1} n_T^{BE}(E) &=& n^{BE}(E) .
\end{eqnarray}
A derivation of the Tsallis distribution, based on the Boltzmann equation,
has been given in Ref.~\cite{Biro:2005uv}.
A comparison between the $n_T^{FD}$ and $n^{FD}$ distributions   
is shown in Fig.~(\ref{compare-tsallis-fd}).
For the Boltzmann approximation, the Tsallis distribution is 
always larger than the Boltzmann one if $q>1$. Taking 
into account the large $p_T$ results for 
particle production we will only consider this possibility in this paper.
As a consequence, in order to keep
the particle yields the same, the Tsallis distribution always leads to
smaller values of the freeze-out temperature for the same set of 
particle yields~\cite{Cleymans:2008mt}.
The Tsallis distribution for quantum statistics has been considered  
in Ref.~\cite{Teweldeberhan:2002wv,Plastino:2004ge,turkey1,Pennini1995309,Chen200265}.
\section{Thermodynamic Consistency}
%
%
%From first law of thermodynamics, we have this relation
%\begin{equation}\label{a2}
% dE = -PdV + TdS + \mu dN,
%\end{equation}
%which for a homogeneous system becomes
%\begin{equation}\label{a1}
% E = -PV + TS + \mu N,
%\end{equation}
%Combining relations~\ref{a1} and~\ref{a2} one obtains 
The first and second laws of thermodynamics lead to the following two 
differential relations~\cite{deGroot:1980aa}
\begin{eqnarray}
 d\epsilon = Tds + \mu dn,\label{a5}\\
dP = sdT + nd\mu.\label{a51}
\end{eqnarray}
where $\epsilon = E/V$, $s = S/V$ and $n = N/V$.
Since these are total differentials, thermodynamic consistency 
requires that the following relations be satisfied
\begin{eqnarray}\label{a6}
 T &=& \left.\frac{\partial \epsilon}{\partial s}\right|_{n},\label{a61}\\
 \mu &=&\left.\frac{\partial \epsilon}{\partial n}\right|_{s},\\
 n &=& \left.\frac{\partial P}{\partial \mu}\right|_{T},\label{a63}\\
 s &=& \left.\frac{\partial P}{\partial T}\right|_{\mu}.\label{a64}
\end{eqnarray}
The pressure, energy density and entropy density are all given by corresponding
integrals over Tsallis distributions and the derivatives have 
to reproduce the corresponding physical quantities. 
For completeness, in the next section, we derive
Tsallis thermodynamics using the maximal entropy principle
and discuss  quantum 
$q$-statistics in particular Bose-Einstein and Fermi-Dirac distribution by maximizing the entropy of the system for quantum
 distributions. This follows partly the derivation of Ref.~\cite{Conroy:2010wt}.
We will show that the consistency conditions given above are indeed 
obeyed by the Tsallis-FD distribution.
%compare the difference between BG and generalized Tsallis thermodynamics which is discussed in this chapter.
%
%
%
\section{Quantum Statistics}
The entropy in standard statistical mechanics for fermions is given 
in the large volume limit by:
\begin{eqnarray}\label{genfermi}
S^{FD}=-gV&&\displaystyle\int \frac{d^3p}{(2\pi)^3}
\left[ n^{FD}\ln n^{FD}\right.\nonumber\\
&&+\left.(1-n^{FD}) \ln(1-n^{FD}) \right],
\end{eqnarray}
where $g$ is the degeneracy factor and $V$ the volume of the system. For 
simplicity Eq.~(\ref{genfermi}) refers to one particle species but can 
be easily generalized to many by summing over all of them.
In the limit where momenta are quantized, which is given by:
\begin{equation}\label{genfermi-small}
S^{FD}=-g\displaystyle\sum_{i}\left[ n_i\ln n_i+(1-n_i) \ln(1-n_i) \right],
\end{equation}
For convenience we will work with the discrete form in the rest of this section.
The large volume limit can be recovered with the standard
replacement:
\begin{equation}
\displaystyle\sum_i \rightarrow V\displaystyle\int \frac{d^3p}{(2\pi)^3}
\end{equation}
The generalization, using the Tsallis prescription, leads to  
\cite{turkey1,Pennini1995309,Teweldeberhan:2005wq}
\begin{equation}\label{genfermi-tsallis}
S^{FD}_T=-g\displaystyle\sum_{i}\left[ n_{i}^{q}\ln_{q} n_{i}+(1-n_{i})^{q} \ln_{q}(1-n_{i}) \right],
\end{equation}
where use has been made of the function 
\begin{equation}
\ln_q (x)\equiv \frac{x^{1-q}-1}{1-q} , \label{suba} 
\end{equation}
often referred to as q-logarithm. 
It can be easily shown that in the limit where the Tsallis parameter $q$
tends to 1 one has:
\begin{equation}
\lim_{q\rightarrow 1}\ln_q (x) = \ln (x) . 
\end{equation} 
The maximization of the entropy~\eqref{genfermi-tsallis} will 
give the  $n_i$'s their  Tsallis-type form.
If we use the explicit form of the  “q-logarithms” we obtain
\begin{equation}\label{FD}
S^{FD}_T= g \displaystyle\sum_{i}\left[ \left( \frac{n_{i}-n_{i}^{q}}{q-1}\right) +\left( \frac{(1-n_{i})-(1-n_{i})^{q}}{q-1}\right) \right],
\end{equation}
In a similar vein,  the generalized form of the  entropy 
for bosons is given by
\begin{equation}\label{genboson-tsallis}
S^{BE}_T=- g\displaystyle\sum_i\left[ n_i^q\ln_{q} n_i-(1+n_i)^q \ln_{q}(1+n_i) \right],
\end{equation}
by using a similar method, we can express 
equation \eqref{genboson-tsallis} as
\begin{equation}\label{B}
S^{BE}_T= g\displaystyle\sum_{i}\left[ \left( \frac{n_{i}-n_{i}^{q}}{q-1}\right) -\left( \frac{(1+n_{i})-(1+n_{i})^{q}}{q-1}\right) \right],
\end{equation}
In the limit $q\rightarrow 1$ equations \eqref{genfermi-tsallis} and 
\eqref{genboson-tsallis} reduce to the standard Fermi-Dirac and 
Bose-Einstein distributions.
Further, as we shall presently explain, the formulation 
of a variational principle in terms of equation \eqref{FD} allows 
to prove the general relation of thermodynamics. 
One of the relevant constraints 
is given by the average number of particles,
\begin{equation}\label{n1}
%\displaystyle\sum_{i} n_{i}^{q} = \left \langle N \right \rangle_{q}.
\displaystyle\sum_{i} n_{i}^{q} =  N .
\end{equation}
Notice the unusual power of $q$ on the left-hand side. As it turns out, 
it is necessary to have this power of $q$ since otherwise there is no 
thermodynamic consistency.

Likewise, the energy of the system gives a constraint,
\begin{equation}\label{n2}
%\displaystyle\sum_{i} n_{i}^{q}E_{i}=\left \langle E \right \rangle_{q}.
\displaystyle\sum_{i} n_{i}^{q}E_{i} = E .
\end{equation}
again, it is necessary to have the power $q$ on the left-hand side as no 
thermodynamic consistency would be achieved without it.
The maximization of the entropic measure 
equation \eqref{FD} under the constraints 
equation \eqref{n1} and \eqref{n2} leads
to the variational problem.
\begin{equation}\label{n3}
\frac{\delta}{\delta n_{i}}
%\left[ S^{FD}_T+\alpha(\left \langle N \right \rangle_{q}-\displaystyle\sum_{i} n_{i}^{q}) +\beta(\left \langle E \right \rangle_{q}-\displaystyle\sum_{i} n_{i}^{q}E_{i})\right] =0,
\left[ S^{FD}_T+\alpha (N -\displaystyle\sum_i n_i^q) +\beta(E -\displaystyle\sum_i n_i^qE_i)\right] =0,
\end{equation}
where $\alpha$ and $\beta$ are Lagrange multipliers 
associated, respectively, with the total number of particles and 
the total energy. Differentiating each 
expression in equation \eqref{n3}
\begin{equation}\label{n4}
%\frac{\delta}{\delta n_{i}}\left(  S^{FD}_T\right) 
%=\frac{q}{q-1}\left[\left( \frac{1-n_i}{n_i}\right)^{q-1}-1 \right] n_i^{q-1},
\frac{\delta}{\delta n_{i}}\left(  S^{FD}_T\right) 
=\frac{q}{q-1}\left[\left( \frac{1-n_{i}}{n_{i}}\right)^{q-1}-1 \right] n_{i}^{q-1},
\end{equation}
\begin{equation}\label{n5}
%\frac{\delta}{\delta n_{i}}\left(\left \langle N \right \rangle_{q}- 
%\displaystyle\sum_{i}n_{i}^{q}\right) = -qn_{i}^{q-1},
\frac{\delta}{\delta n_{i}}\left( N - \displaystyle\sum_in_i^q\right) 
= -qn_i^{q-1},
\end{equation}
and
\begin{equation}\label{n6}
%\frac{\delta}{\delta n_{i}}\left(\left \langle E \right \rangle_{q}-\displaystyle\sum_{i}n_{i}^{q}E_{i}\right)
\frac{\delta}{\delta n_{i}}\left(E - \displaystyle\sum_in_i^qE_i\right)
 =-qE_in_i^{q-1},
\end{equation}
then by substituting equation \eqref{n4}, \eqref{n5} and 
\eqref{n6} into \eqref{n3}, we obtain
\begin{equation}\label{n7}
qn_{i}^{q-1} \left\lbrace \frac{1}{q-1}\left[  -1+\left(\frac{1-n_{i}}{n_{i}} \right)^{q-1} \right]
-\beta E_{i}-\alpha \right\rbrace=0.
\end{equation}
which can be rewritten as
\begin{equation}\label{n8}
\frac{1}{q-1}\left[  -1+\left(\frac{1-n_{i}}{n_{i}} \right)^{q-1} \right]=\beta E_{i}+\alpha,
\end{equation}
and, by rearranging equation \eqref{n8}, we get
\begin{displaymath}
\frac{1-n_{i}}{n_{i}}=\left[ 1+(q-1)(\beta E_{i}+\alpha)\right]^{\frac{1}{q-1}},
\end{displaymath}
finally, we get the solution of generalized form of 
Fermi-Dirac distribution like this
\begin{eqnarray}
n_{i}&=&\frac{1}{\left[ 1+(q-1)(\beta E_{i}+\alpha)\right]^{\frac{1}{q-1}}+1},\nonumber \\
     &=&\frac{1}{\left[\exp_{q}(\alpha +\beta E_{i} )\right] +1},
\end{eqnarray}
Which is the expression for  the Tsallis-FD distribution referred to earlier
in this paper~\cite{turkey1,Pennini1995309,Teweldeberhan:2005wq}.

Using a similar approach one can also determine the Tsallis-BE 
distribution~\cite{Chen200265}. 
Starting from the extremum of the entropy subject to two conditions one has:
\begin{equation}\label{n10}
\frac{\delta}{\delta n_{i}}\left[ S^{BE}_T+\alpha(N -\displaystyle\sum_{i} 
n_i^{q}) +\beta(E -\displaystyle\sum_{i} n_{i}^{q}E_{i})\right] =0,
\end{equation}
which leads to
\begin{equation}\label{n11}
\frac{\delta}{\delta n_{i}}\left(  S^{BE}_T\right)=
\frac{q}{q-1}\left[\left( \frac{1+n_{i}}{n_{i}}\right)^{q-1}-1 \right] n_{i}^{q-1},
\end{equation}
and  by using equations \eqref{n11},\eqref{n5} and \eqref{n6} in \eqref{n10}, one gets
\begin{equation}\label{n12}
\displaystyle\sum_{i}qn_{i}^{q-1} \left\lbrace \frac{1}{q-1}\left[  -1+\left(\frac{1+n_{i}}{n_{i}} \right)^{q-1} \right]-\beta E_{i}-\alpha \right\rbrace=0.
\end{equation}
By rearranging equation \eqref{n12},  one obtains the expression  for 
the Tsallis-BE distribution~\cite{Chen200265},

\begin{eqnarray}
n_{i}&=&\frac{1}{\left[ 1+(q-1)(\beta E_{i}+\alpha)\right]^{\frac{1}{q-1}}-1},\nonumber \\
     &=&\frac{1}{\left[\exp_{q}((E_{i} -\mu)/T)\right]-1}  .\label{q}
\end{eqnarray}
where the usual identifications $\alpha = -\mu/T$  and $\beta = 1/T$ have been made. 
\section{Proof of Thermodynamical Consistency}
In order to use the above expressions it has to be shown that they satisfy 
the thermodynamic consistency conditions. To show this in detail 
we use the first law of thermodynamics~\cite{deGroot:1980aa}
\begin{equation}\label{a10}
 P = \frac{-E + TS + \mu N}{V}, 
\end{equation}
and take the partial derivative with respect to $\mu$ in
order to check for thermodynamic consistency, it leads to
\begin{eqnarray}\label{a11}
\left.\frac{\partial P}{\partial \mu}\right|_T & = &
\frac{1}{V}
\left[-\frac{\partial E}{\partial \mu} +T\frac{\partial S}{\partial \mu} + N + \mu\frac{\partial N}{\partial \mu}\right],\nonumber \\  
% & = &\frac{1}{V}\left[N + \displaystyle\sum_{i}\left(\mu - E_{i} - 
% \frac{T}{q-1}\right)\frac{\partial n_{i}^{q}}{\partial \mu} + \frac{Tq(1-n_{i})^{q-1}}{q-1}\frac{\partialn_{i}}{\partial \mu}
% \right],\nonumber \\
% & = &\frac{1}{V}\left[N + \displaystyle\sum_{i}-\frac{T}{q-1}\left(1 + 
% \frac{(q-1)(E_{i} -\mu)}{T}\right)\frac{\partial n_{i}^{q}}{\partial \mu} + 
% \frac{Tq(1-n_{i})^{q-1}}{q-1}\frac{\partialn_{i}}{\partial \mu}
% \right],\nonumber \\
& = &\frac{1}{V}\left[N + \displaystyle\sum_{i}-\frac{T}{q-1}\left(1 + 
(q-1)\frac{E_{i} -\mu}{T}\right)\frac{\partial n_{i}^{q}}{\partial \mu} \right.\nonumber \\
&& \left.+\frac{Tq(1-n_{i})^{q-1}}{q-1}\frac{\partial n_{i}}{\partial \mu}
\right],
\end{eqnarray}
then, by explicit calculation
\begin{displaymath}
 \frac{\partial n_{i}^{q}}{\partial \mu} = \frac{qn_{i}^{q+1}}{T}\left[1+(q-1)\frac{E_{i}-\mu}{T}\right]^{-1+\frac{1}{1-q}},
\end{displaymath}
\begin{displaymath}
 \frac{\partial n_{i}}{\partial \mu} = \frac{n_{i}^{2}}{T}\left[1+(q-1)\frac{E_{i}-\mu}{T}\right]^{-1+\frac{1}{1-q}},
\end{displaymath}
and
\begin{displaymath}
\left(1-n_{i}\right)^{q-1} = n_{i}^{q-1}\left[1+\frac{(q-1)(E_{i}-\mu)}{T}\right].
\end{displaymath}
Introducing this  into equation \eqref{a11},  yields
\begin{equation}\label{a12}
\left. \frac{\partial P}{\partial \mu}\right|_T = n,
\end{equation}
which proves the thermodynamical consistency \eqref{a63}. 

We also calculate explicitly the relation in equation \eqref{a61} can be rewritten as
\begin{align}\label{a13}
\left. \frac{\partial E}{\partial S}\right|_n& = \frac{\frac{\partial E}{\partial T}dT + \frac{\partial E}{\partial \mu}d\mu}{\frac{\partial S}{\partial T}dT
 + \frac{\partial S}{\partial \mu}d\mu},\nonumber \\
& = \frac{\frac{\partial E}{\partial T} + \frac{\partial E}{\partial \mu}\frac{d\mu}{dT}}{\frac{\partial S}{\partial T}
 + \frac{\partial S}{\partial \mu}\frac{d\mu}{dT}},
\end{align}
since $n$ is kept fixed one has the additional constraint
\begin{displaymath}
 dn = \frac{\partial n}{\partial T}dT + \frac{\partial n}{\partial \mu}d\mu = 0,
\end{displaymath}
leading to
\begin{equation}\label{a18}
 \frac{d\mu}{dT} = -\frac{\frac{\partial n}{\partial T}}{\frac{\partial n}{\partial \mu}}.
\end{equation}
Now, we rewrite \eqref{a13} and \eqref{a18} in terms of the following 
expressions
\begin{displaymath}
 \frac{\partial E}{\partial T} = \displaystyle\sum_{i} qE_{i}n_{i}^{q-1}\frac{\partial n_{i}}{\partial T}, 
\end{displaymath}
\begin{displaymath}
 \frac{\partial E}{\partial \mu} = \displaystyle\sum_{i} qE_{i}n_{i}^{q-1}\frac{\partial n_{i}}{\partial \mu}, 
\end{displaymath}
\begin{displaymath}
 \frac{\partial S}{\partial T} = \displaystyle\sum_{i} q\left[\frac{-n_{i}^{q-1}+(1-n_{i})^{q-1}}{q-1}\right]\frac{\partial n_{i}}{\partial T}, 
\end{displaymath}
\begin{displaymath}
 \frac{\partial S}{\partial \mu} = \displaystyle\sum_{i}q\left[\frac{-n_{i}^{q-1}+(1-n_{i})^{q-1}}{q-1}\right]\frac{\partial n_{i}}{\partial \mu}, 
\end{displaymath}
\begin{displaymath}
 \frac{\partial n}{\partial T} = \frac{1}{V}\left[\displaystyle\sum_{i} qn_{i}^{q-1}\frac{\partial n_{i}}{\partial T}\right], 
\end{displaymath}
and
\begin{displaymath}
 \frac{\partial n}{\partial \mu} = \frac{1}{V}\left[\displaystyle\sum_{i} qn_{i}^{q-1}\frac{\partial n_{i}}{\partial \mu}\right]. 
\end{displaymath}
By introducing the above relations into equation \eqref{a13}, the numerator 
 of equation \eqref{a13} becomes
\begin{eqnarray}\label{a19}
\frac{\partial E}{\partial T} &+& \frac{\partial E}{\partial \mu}\frac{d\mu}{dT} 
 = \displaystyle\sum_{i} qE_{i}n_{i}^{q-1}\frac{\partial n_i}{\partial T}\nonumber \\
&&-\frac{\displaystyle\sum_{i,j} q^{2}E_{j}\left(n_{i}n_{j}\right)^{q-1}\frac{\partial n_{j}}{\partial \mu}\frac{\partial n_{i}}{\partial T}}{\displaystyle\sum_{j} 
qn_{j}^{q-1}\frac{\partial n_{j}}{\partial \mu}},\nonumber \\ 
& =& \frac{\displaystyle\sum_{i,j} qE_{i}\left(n_{i}n_{j}\right)^{q-1}C_{ij}}
{\displaystyle\sum_{j} 
n_{j}^{q-1}\frac{\partial n_{j}}{\partial \mu}}.
\end{eqnarray}
Where the abbreviation 
\begin{equation}
 C_{ij}\equiv \left(n_{i}n_{j}\right)^{q-1}\left[\frac{\partial n_{i}}{\partial T}\frac{\partial n_{j}}{\partial \mu}-
 \frac{\partial n_{j}}{\partial T}\frac{\partial n_{i}}{\partial \mu}\right],
\end{equation}
has been introduced. 
One can rewrite the denominator part of equation \eqref{a13}  as
\begin{align}\label{a20}
\frac{\partial S}{\partial T} + \frac{\partial S}{\partial \mu}\frac{d\mu}{dT} & = 
% \displaystyle\sum_{i} q\left[\frac{-n_{i}^{q-1}+(1-n_{i})^{q-1}}{q-1}\right]\frac{\partial n_{i}}{\partial T}
% -\displaystyle\sum_{j} q\left[\frac{-n_{j}^{q-1}+(1-n_{j})^{q-1}}{q-1}\right]\frac{\partial n_{j}}{\partial \mu}
% \frac{\displaystyle\sum_{i} qn_{i}^{q-1}\frac{\partial n_{i}}{\partial T}}{\displaystyle\sum_{j} 
% qn_{j}^{q-1}\frac{\partial n_{j}}{\partial \mu}},\nonumber \\ 
 \frac{\displaystyle q\sum_{i,j}\left[{-n_{i}^{q-1}+(1-n_{i})^{q-1}}\right]n_{j}^{q-1}
 C_{i,j}}
 {(q-1)\displaystyle\sum_{j} 
n_{j}^{q-1}\frac{\partial n_{j}}{\partial \mu}},\nonumber \\
& = \frac{\displaystyle q\sum_{i,j}(E_{i}-\mu)\left(n_{i}n_{j}\right)^{q-1}
 C_{i,j}}
 {T\displaystyle \sum_{j} 
n_{j}^{q-1}\frac{\partial n_{j}}{\partial \mu}},
\end{align}
where
\begin{displaymath}
 \frac{-n_{i}^{q-1}+(1-n_{i})^{q-1}}{q-1} = \frac{(E_{i}-\mu)}{T}n_{i}^{q-1},
\end{displaymath}
hence, by substituting equation \eqref{a19} and \eqref{a20} in to \eqref{a13}, we find
\begin{equation}\label{a21}
\left. \frac{\partial E}{\partial S}\right|_n = T\frac{\displaystyle\sum_{i,j}E_{i}C_{ij}}
{\displaystyle\sum_{i,j}(E_{i}-\mu)C_{ij}},
\end{equation}
since $\displaystyle\sum_{i,j} C_{ij} = 0$, this finally leads to the desired result
\begin{equation}
\left.\frac{\partial E}{\partial S}\right|_n = T.
\end{equation}
Hence thermodynamic consistency is satisfied.
 
\section{Boltzmann Approximation}
%\section{The modified Tsallis Entropy}
Due to its practical relevance and importance we devote a section to the 
Tsallis-B distribution. In this case the entropy 
is obtained from equation \eqref{a64} 
by assuming the  $n_i \ll 1$, this leads to
\begin{equation}\label{a14}
 S_T^B\equiv g\displaystyle\sum_{i=1}^{W}\frac{(n_{i}-n_{i}^{q})}{q-1}+n_{i},
\end{equation}
The $n_i$  are given explicitly as
\begin{equation}\label{a15}
 {n}_{i}=\left[ 1+(q-1)\frac{E_{i}-\mu}{T}\right]^{\frac{1}{1-q}},
\end{equation}
where $n_{i}$ denotes the number of particles in the $i$th energy 
level with energy $E_i$. The
maximum of the above entropy  is looked for 
 under the constraints imposed by fixing the total
number of particles $N$ and the total energy of the system $E$, as given 
in equation \eqref{n1} and \eqref{n2}. 
As in the previous section, it should satisfy  thermodynamic 
consistency which is given in equation
\eqref{a6}. The derivative of pressure w.r.t. $\mu$ becomes
\begin{align}\label{a16}
\left.\frac{\partial P}{\partial \mu}\right|_T& = \frac{1}{V}\left[-\frac{\partial E}{\partial \mu} +T\frac{\partial S}{\partial \mu} + N + \mu\frac{\partial N}{\partial \mu}\right],\nonumber \\  
% & = \frac{1}{V}\left[N + \displaystyle\sum_{i}\left(-(E_{i}-\mu) - 
% \frac{T}{q-1}\right)\frac{\partial n_{i}^{q}}{\partial \mu} + \frac{T}{q-1}\frac{\partialn_{i}}{\partial \mu} 
% + T\frac{\partialn_{i}}{\partial \mu}
% \right],\nonumber \\
% & = \frac{1}{V}\left[N + \displaystyle\sum_{i}-\frac{T}{q-1}\left(1 + 
% \frac{(q-1)(E_{i} -\mu)}{T}\right)\frac{\partial n_{i}^{q}}{\partial \mu} + 
% T\left(\frac{1}{q-1}+1\right)\frac{\partialn_{i}}{\partial \mu}
% \right],\nonumber \\
& = \frac{1}{V}\left[N + \displaystyle\sum_{i}-\frac{Tn_{i}^{1-q}}{q-1}\frac{\partial n_{i}^{q}}{\partial \mu} + 
\frac{Tq}{q-1}\frac{\partial n_{i}}{\partial \mu},
\right], 
\end{align}
now, by using
\begin{displaymath}
 \frac{\partial n_{i}^{q}}{\partial \mu} = qn_{i}^{q-1}\frac{\partial n_{i}}{\partial \mu},
\end{displaymath}
and
\begin{displaymath}
 \frac{\partial n_{i}}{\partial \mu} = \frac{n_{i}^{q}}{T} .
\end{displaymath}
By   the above relations in equation \eqref{a16}, we 
recover equation \eqref{a63}.

We  now calculate the expressions needed in 
equations \eqref{a13} and \eqref{a18} in terms of
\begin{displaymath}
 \frac{\partial S}{\partial T} = \displaystyle\sum_{i} \left[1+\frac{1-qn_{i}^{q-1}}{q-1}\right]\frac{\partial n_{i}}{\partial T}, 
\end{displaymath}
\begin{displaymath}
 \frac{\partial S}{\partial \mu} = \displaystyle\sum_{i} \left[1+\frac{1-qn_{i}^{q-1}}{q-1}\right]\frac{\partial n_{i}}{\partial \mu}, 
\end{displaymath}
while the other partial derivatives are the same as previously.
by plugging the above relations into equation \eqref{a13}, then the numerator part of equation \eqref{a13} become
%by plugging these relations to equation \eqref{a13}, the numerator part become 
\begin{eqnarray}\label{a199}
\frac{\partial E}{\partial T} + 
\frac{\partial E}{\partial \mu}\frac{d\mu}{dT} 
& = &\displaystyle\sum_{i} qE_{i}n_{i}^{q-1}
\frac{\partial n_{i}}{\partial T} \nonumber \\
&& -\frac{\displaystyle\sum_{i,j} q^{2}E_{j}\left(n_{i}n_{j}\right)^{q-1}\frac{\partial n_{j}}{\partial \mu} 
\frac{\partial n_{i}}{\partial T}}{\displaystyle\sum_{j} 
qn_{j}^{q-1}\frac{\partial n_{j}}{\partial \mu}},\nonumber \\ 
& = &\frac{\displaystyle\sum_{i,j} qE_{i}\left(n_{i}n_{j}\right)^{q-1}C_{i,j}}
{\displaystyle\sum_{j} 
n_{j}^{q-1}\frac{\partial n_{j}}{\partial \mu}} .
\end{eqnarray}
Similarly, the denominator part of equation \eqref{a13} can be  written as
\begin{eqnarray}\label{a200}
\frac{\partial S}{\partial T} &+& \frac{\partial S}{\partial \mu}\frac{d\mu}{dT}  = 
\displaystyle\sum_{i} \left[1+\frac{1-qn_{i}^{q-1}}{q-1}\right]\frac{\partial n_{i}}{\partial T}\nonumber \\
&& -
\frac{\displaystyle\sum_{i,j} n_{i}^{q-1}\left[1+\frac{1-qn_{j}^{q-1}}{q-1}\right]\frac{\partial n_{j}}{\partial \mu}
 \frac{\partial n_{i}}{\partial T}}{\displaystyle\sum_{j} 
n_{j}^{q-1}\frac{\partial n_{j}}{\partial \mu}},\nonumber \\ 
 & = &\frac{\displaystyle\sum_{i,j}\left[1+\frac{1-qn_{i}^{q-1}}{q-1}\right]n_{j}^{q-1}
 C_{i,j}}
 {\displaystyle\sum_{j} 
n_{j}^{q-1}\frac{\partial n_{j}}{\partial \mu}},\nonumber \\
& = &\frac{\displaystyle\sum_{i,j}\frac{q(E_{i}-\mu)}{T}\left(n_{i}n_{j}\right)^{q-1}
 C_{i,j}}
 {\displaystyle\sum_{j} 
n_{j}^{q-1}\frac{\partial n_{j}}{\partial \mu}},
\end{eqnarray}
where
\begin{displaymath}
 1+\frac{1-qn_{i}^{q-1}}{q-1} = \frac{q(E_{i}-\mu)}{T}n_{i}^{q-1},
\end{displaymath}
by combining the expressions in 
equation \eqref{a199} and \eqref{a200} into \eqref{a13}, we find as before
\begin{equation}\label{a22}
\left.\frac{\partial E}{\partial S}\right|_n = T .
\end{equation}
It has thus been shown that the definitions of temperature and
pressure within the Tsallis formalism 
for non-extensive thermostatistics lead to expressions which
satisfy consistency with the first law of thermodynamics.

%%%%%%%%%%%%%%%%%%%%%%%%%%%%%%%%%%%%%%%%%%%%%%%%%%%%%%%%%%%%%%%%%%%
%
%
%
\section{Thermal Fit Details} %%%%%%%%%%%%%%%%%%%%%%%%%%%%%%%%%%%%%%%%%%
The total number of particles is given by the  integral version of~\eqref{n1}, 
\begin{equation}
N = gV\displaystyle\int \frac{d^3p}{(2\pi)^3}
\left[1+(q-1)\frac{E-\mu}{T}\right]^{q/(1-q)},
\end{equation}
The extra power of $q$ is necessary for thermodynamic consistency.
The corresponding (invariant)  momentum distribution is given by 
\begin{equation}
E\frac{dN}{d^3p} = gVE\frac{1}{(2\pi)^3}
\left[1+(q-1)\frac{E-\mu}{T}\right]^{q/(1-q)},
\end{equation}
which, in terms of the rapidity and transverse mass variables, becomes

\begin{eqnarray}
\frac{dN}{dy\, m_Tdm_T} &=& gV\frac{m_T\cosh y}{(2\pi)^2}\nonumber\\
&&\times\left[1+(q-1)\frac{m_T\cosh y -\mu}{T}\right]^{q/(1-q)},
\end{eqnarray}
At mid-rapidity $y=0$ and for zero chemical potential this reduces to
the following expression
\begin{equation}\label{alice}
\left.\frac{dN}{m_T dm_T~dy}\right|_{y=0} = gV\frac{m_T}{(2\pi)^2}
\left[1+(q-1)\frac{m_T}{T}\right]^{q/(1-q)},
\end{equation}
or, introducing  the transverse momentum:
\begin{equation}
\left.\frac{dN}{dp_T~dy}\right|_{y=0} = gV\frac{p_Tm_T}{(2\pi)^2}
\left[1+(q-1)\frac{m_T}{T}\right]^{q/(1-q)}.
\end{equation}
Fits using the above expressions based on the Tsallis-B distribution
 to experimental 
measurements published by the CMS collaboration~\cite{Khachatryan:2011tm}
are shown in Figs.~2, 3 and 4 and are comparable with those 
shown by the CMS  collaboration
but the resulting parameters are considerably different and are collected 
in Table I.  The most striking feature is that the values of 
the parameter $q$ are fairly stable around the value $q \approx 1.11$ for all cases 
considered, whether 7 Tev or 0.9 TeV. The same cannot be said about the 
temperature $T$ which is around 100 MeV with considerable deviations, it
is however well below the values quoted by the 
CMS collaboration~\cite{Khachatryan:2011tm}.
%
%The volume factor $V$ cannot be interpreted as the volume of the system
%as one expects contributions from the decays of resonances.
%
The analytic expression used in Refs.~\cite{Aamodt:2011zj,Khachatryan:2011tm} 
corresponds to 
identifying 
\begin{equation}
n\rightarrow \frac{q}{q-1}
\end{equation}
an additional factor of the transverse mass on the right-hand side and
a shift in the mass.
\begin{table}[ht]
\begin{center}
\begin{tabular}{|c|c|c|}
\hline
Particle             & $T$ (MeV) & $q$     \\
\hline 
$K^0_S$  (0.9 TeV)     & 92    & 1.13 \\
$K^0_S$  (7 TeV)       & 105   & 1.15  \\
$\Lambda$  (0.9 TeV)   & 70    & 1.11  \\
$\Lambda$  (7 TeV)     & 117   & 1.12  \\
$\Xi^-$  (0.9 TeV)     & 44    & 1.11  \\
$\Xi^-$  (7 TeV)       & 126   & 1.11  \\
\hline  
\end{tabular}
\caption{Fitted values of the $T$ and $q$ parameters for strange particles 
measured by
the CMS collaboration~\cite{Khachatryan:2011tm} using the Tsallis-B form 
for the momentum distribution. The normalization has been adjusted. 
}
\end{center}
\end{table}
\section{Conclusions}
In this paper we have presented a detailed derivation  of the quantum form
of the Tsallis distribution and considered in detail 
the thermodynamic consistency of the resulting distribution.
It was emphasized that an additional power of $q$ is needed to achieve
consistency with the laws of thermodynamics~\cite{Conroy:2010wt}.
The resulting distribution, called Tsallis-B, was compared with 
recent measurements from the CMS collaboration~\cite{Khachatryan:2011tm}
and good agreement was obtained. The resulting parameter $q$ which is
a measure for the deviation from a standard Boltzmann distribution was found
to be around 1.11.  The resulting values of the temperature show a wider spread
around 100 MeV.

Whether or not the Tsallis distribution  provides a valid 
interpretation of high energy collision data will need further theoretical work.
\bibliography{cleymans_worku}
\newpage
\begin{figure}
\begin{center}
\includegraphics[width=0.5\textwidth,height=10cm]{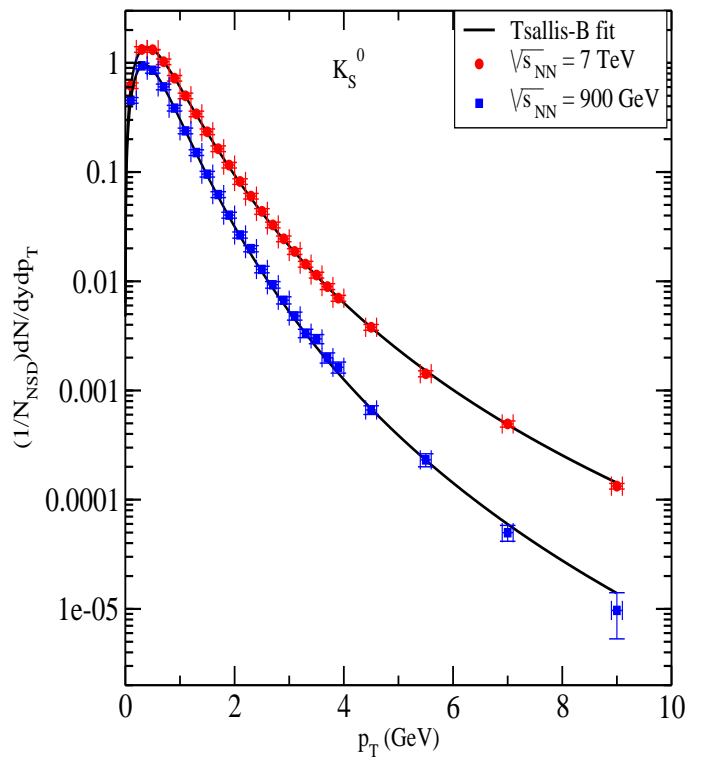}
\caption{Comparison between the 
measured transverse momentum distribution for $K^0_S$ as measured by 
the CMS collaboration~\cite{Khachatryan:2011tm} and  the Tsallis-B distribution
as given by Eq.~\eqref{alice} using the parameters listed in Table I.}
\label{CMS-K0S}
\end{center}
\end{figure}
\begin{figure}
\begin{center}
\includegraphics[width=0.5\textwidth,height=10cm]{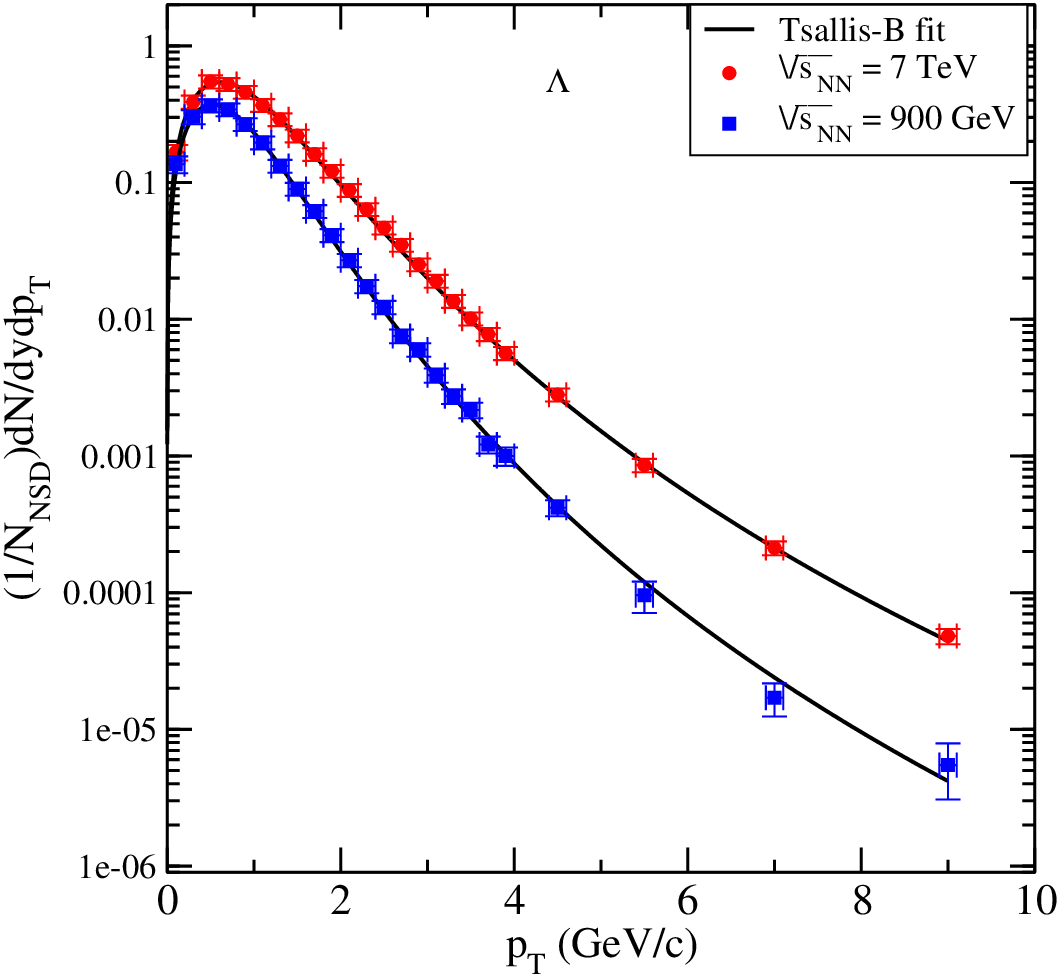}
\caption{Comparison between the 
measured transverse momentum distribution for $\Lambda$ as measured by 
the CMS collaboration~\cite{Khachatryan:2011tm} and  the Tsallis-B distribution
as given by Eq.~\eqref{alice} using the parameters listed in Table I.}
\label{CMS-Lambda}
\end{center}
\end{figure}
\begin{figure}
\begin{center}
\includegraphics[width=0.5\textwidth,height=10cm]{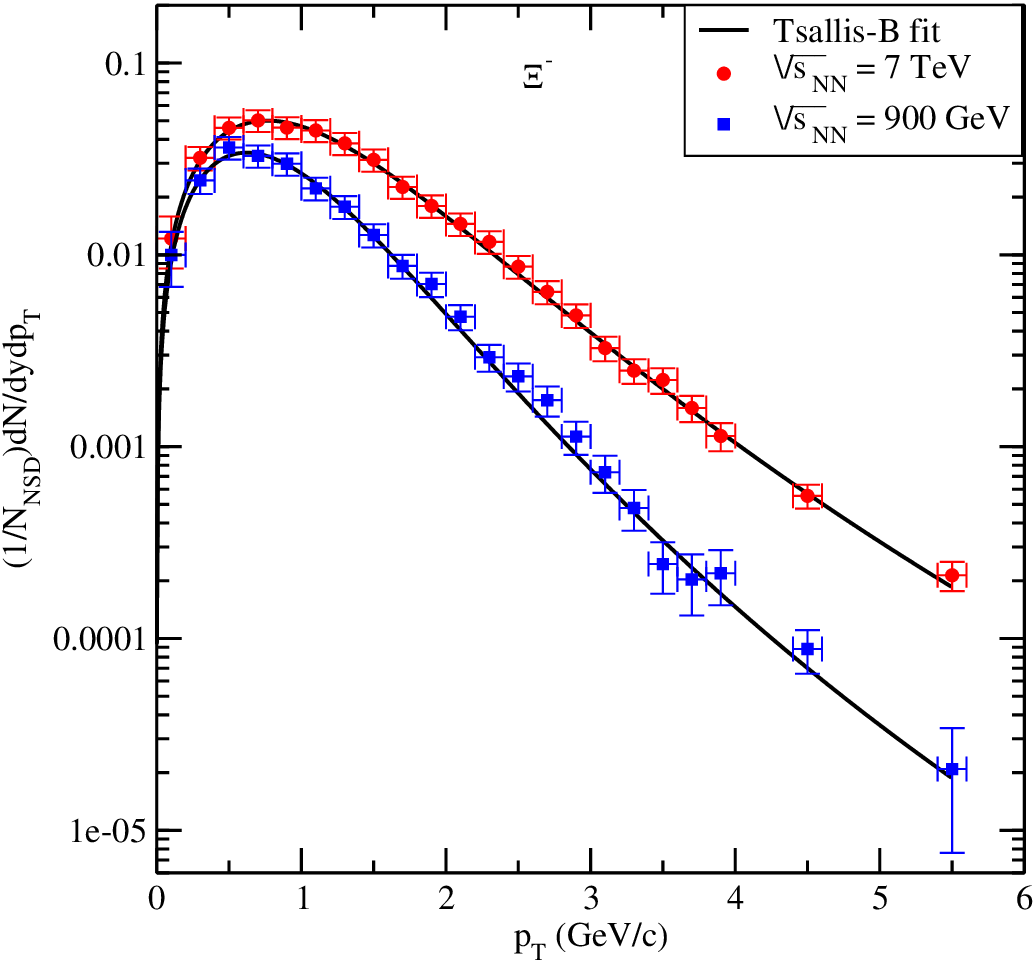}
\caption{Comparison between the 
measured transverse momentum distribution for $\Lambda$ as measured by 
the CMS collaboration~\cite{Khachatryan:2011tm} and  the Tsallis-B distribution
as given by Eq.~\eqref{alice} using the parameters listed in Table I.}
\label{CMS-Xi}
\end{center}
\end{figure}
\end{document}